\documentclass[fleqn,usenatbib]{mnras}

\usepackage{newtxtext,newtxmath}
\usepackage[T1]{fontenc}

\DeclareRobustCommand{\VAN}[3]{#2}
\let\VANthebibliography\thebibliography
\def\thebibliography{\DeclareRobustCommand{\VAN}[3]{##3}\VANthebibliography}

\usepackage{pdflscape}
\usepackage{rotating}
\usepackage{caption}
\usepackage{enumitem}
\usepackage{graphicx}
\graphicspath{./figs/}
\usepackage{amsmath}
\usepackage{natbib}
\usepackage{bm}
\usepackage{color}
\usepackage{natbib}
\usepackage{amsfonts}
\usepackage{multirow}
\usepackage{aas_macros}
\usepackage{threeparttable}
\usepackage{url}
\usepackage{xcolor}
\usepackage{twoopt}
\usepackage{orcidlink}
\usepackage{subcaption}
\usepackage{mathtools}

\newcommandtwoopt{\myfrac}[4][0pt][0pt]{\genfrac{}{}{}{}{\raisebox{#1}{$#3$}}{\raisebox{-#2}{$#4$}}}
\title[UV suppression impact on LSST lensed SNIIn]{The impact of ultraviolet suppression on the rates and properties of strongly lensed Type IIn supernovae detected by LSST}

\author[Andr\'es I. Ponte P\'erez et al.]{Andr\'es\ I.\ Ponte\         P\'erez,\!$^{1}$\thanks{E-mail: aip981@student.bham.ac.uk}\orcidlink{},
Graham\ P.\ Smith,\!$^{1,2}$\orcidlink{0000-0003-4494-8277}
Matt\ Nicholl,\!$^{3}$\orcidlink{0000-0002-2555-3192}
Nikki\ Arendse,\!$^{4}$\orcidlink{0000-0001-5409-6480}
Dan\ Ryczanowski,\!$^{5,1}$\orcidlink{0000-0002-4429-3429} \newauthor
Suhail\ Dhawan$^{1}$ and
the LSST Strong Lensing Science Collaboration
\\
$^{1}$School of Physics and Astronomy, University of Birmingham, Edgbaston, B15 2TT, UK\\
$^{2}$Department of Astrophysics, University of Vienna, T\"urkenschanzstrasse 17, 1180 Vienna, Austria\\
$^{3}$Astrophysics Research Centre, School of Mathematics and Physics, Queen's University Belfast, Belfast BT7 1NN, UK\\
$^{4}$Oskar Klein Centre, Department of Physics, Stockholm University, SE-106 91 Stockholm, Sweden\\
$^{5}$Institute of Cosmology and Gravitation, University of Portsmouth, Burnaby Road, Portsmouth, PO1 3FX, UK
}

\date{Accepted XXX. Received XXX; in original form XXX.}
\pubyear{\the\year{}}

\begin{document}
\label{firstpage}
\pagerange{\pageref{firstpage}--\pageref{lastpage}}
\maketitle

\newcommand{\red}{\textcolor{red}}
\newcommand{\blue}{\textcolor{blue}}
\newcommand{\magenta}{\textcolor{magenta}}
\newcommand{\norv}[1]{{\textcolor{blue}{#1}}}
\newcommand{\norvnew}[1]{{\textcolor{red}{#1}}}
\newcommand\hkpc{\ensuremath{h^{-1}~\mathrm{kpc}}}
\newcommand\hMpc{\ensuremath{h^{-1}~\mathrm{Mpc}}}
\newcommand\Mpc{\ensuremath\,\mathrm{Mpc}}
\newcommand\Mvir{\ensuremath{M_\mathrm{vir}}}
\newcommand\cvir{\ensuremath{c_\mathrm{vir}}}
\newcommand\rvir{\ensuremath{r_\mathrm{vir}}}
\newcommand\Dvir{\ensuremath{\Delta_\mathrm{vir}}}
\newcommand\rsc{\ensuremath{r_\mathrm{sc}}}
\newcommand\rhoc{\ensuremath{\rho_\mathrm{crit}}}
\newcommand\Msol{\ensuremath\mathrm{M_\odot}}
\newcommand\hMsol{\ensuremath{h^{-1}\mathrm{M_\odot}}}
\newcommand\ergs{\ensuremath\mathrm{erg\,s^{-1}}}
\newcommand\Mgas{\ensuremath{M_\mathrm{gas}}}
\newcommand\Mhse{\ensuremath{M_\mathrm{HSE}}}
\newcommand\Mp{\ensuremath{M_\mathrm{Planck}}}
\newcommand\Mwl{\ensuremath{M_\mathrm{WL}}}
\newcommand\Mfh{\ensuremath{M_{500}}}
\newcommand\rfh{\ensuremath{r_{500}}}
\newcommand\Tx{\ensuremath{T_X}}
\newcommand\Om{\ensuremath{\Omega_\mathrm{M}}}
\newcommand\Ol{\ensuremath{\Omega_\Lambda}}
\newcommand\eV{\,\ensuremath\mathrm{eV}}
\newcommand\keV{\,\ensuremath\mathrm{keV}}
\newcommand\MeV{\,\ensuremath\mathrm{MeV}}
\newcommand\GeV{\,\ensuremath\mathrm{GeV}}
\newcommand\kpc{\,\ensuremath\mathrm{kpc}}
\newcommand\kms{\,\ensuremath\mathrm{km\,s^{-1}}}
\newcommand\m{\,\ensuremath\mathrm{m}}
\newcommand\persqmpersec{\,\ensuremath\mathrm{m^{-2}\,s^{-1}}}
\newcommand\cm{\,\ensuremath\mathrm{cm}}
\newcommand\Watt{\,\ensuremath\mathrm{W}}
\newcommand\ls{\,\ensuremath{\hbox{\rlap{\hbox{\lower4pt\hbox{$\sim$}}}\hbox{$<$}}}\,}
\newcommand\gs{\,\ensuremath{\hbox{\rlap{\hbox{\lower4pt\hbox{$\sim$}}}\hbox{$>$}}}\,}
\newcommand\dl{\ensuremath{D_\mathrm{L}}}
\newcommand\dlf{\ensuremath{D_\mathrm{L,f}}}
\newcommand\ds{\ensuremath{D_\mathrm{S}}}
\newcommand\dls{\ensuremath{D_\mathrm{LS}}}
\newcommand\dlC{\ensuremath{D_\mathrm{L}^{\mathrm C}}}
\newcommand\dsC{\ensuremath{D_\mathrm{S}^{\mathrm C}}}
\newcommand\dlsC{\ensuremath{D_\mathrm{LS}^{\mathrm C}}}
\newcommand\dsi{\ensuremath{D_{\mathrm{S},i}}}
\newcommand\dlsi{\ensuremath{D_{\mathrm{LS},i}}}
\newcommand\dsj{\ensuremath{D_{\mathrm{S},j}}}
\newcommand\dlsj{\ensuremath{D_{\mathrm{LS},j}}}
\newcommand\zs{\ensuremath{z_{\rm S}}}
\newcommand\zl{\ensuremath{z_{\rm L}}}
\newcommand\ks{\ensuremath\mathrm{ks}}
\newcommand\betaP{\ensuremath{\beta_\mathrm{P}}}
\newcommand\betaX{\ensuremath{\beta_\mathrm{X}}}
\newcommand\valpha{\ensuremath{\bm{\alpha}}}
\newcommand\vbeta{\ensuremath{\bm{\beta}}}
\newcommand\vtheta{\ensuremath{\bm{\theta}}}
\newcommand\vchi{\ensuremath{\bm{\chi}}}
\newcommand\vlum{\ensuremath{\bm{\Lambda}}}
\newcommand\mutot{\ensuremath{\mu_{\rm tot}}}
\newcommand\mupeak{\ensuremath{\mu_{\rm peak}}}
\newcommand\mup{\ensuremath{\mu_{\rm p}}}
\newcommand\thE{\ensuremath{\theta_{\rm E}}}
\newcommand\kE{\ensuremath{\kappa_{\rm E}}}
\newcommand\etaE{\ensuremath{\eta_{\rm E}}}
\newcommand\tz{\ensuremath{\widetilde{z}}}
\newcommand\tD{\ensuremath{\widetilde{D}}}
\newcommand\tM{\ensuremath{\widetilde{\mathcal{M}}}}
\newcommand\tm{\ensuremath{\widetilde{m}}}
\newcommand\mMmin{\ensuremath{\mathcal{M}_{\rm min}}}
\newcommand\mMmax{\ensuremath{\mathcal{M}_{\rm max}}}
\newcommand\mmin{\ensuremath{m_{\rm min}}}
\newcommand\mmax{\ensuremath{m_{\rm max}}}
\newcommand\mbr{\ensuremath{m_{\rm br}}}
\newcommand\pdet{\ensuremath{p_{\rm det}}}
\newcommand\mR{\ensuremath{\mathcal{R}}}
\newcommand\mK{\ensuremath{\mathcal{K}}}
\newcommand\mM{\ensuremath{\mathcal{M}}}
\newcommand\mD{\ensuremath{\mathcal{D}}}
\newcommand\Nlens{\ensuremath{N_{\rm lens}}}
\newcommand\Ndet{\ensuremath{N_{\rm det}}}
\newcommand\Rdet{\ensuremath{R_{\rm det}}}
\newcommand\Dmax{\ensuremath{D_{\rm max}}}
\newcommand\zmax{\ensuremath{z_{\rm max}}}
\newcommand\zmin{\ensuremath{z_{\rm min}}}
\newcommand\mlim{\ensuremath{m_{\rm lim}}}
\newcommand\Dr{\ensuremath{D_{\rm R}}}
\newcommand\Narr{\ensuremath{N_{\rm arr}}}
\newcommand\zpivot{\ensuremath{z_{\rm pivot}}}
\newcommand\zpeak{\ensuremath{z_{\rm peak}}}
\newcommand\Dt{\ensuremath{\Delta t}}
\newcommand\Dtheta{\ensuremath{\Delta\theta}}
\newcommand\sigcrit{\ensuremath{\Sigma_{\rm crit}}}
\newcommand\thA{\ensuremath{\theta_{\rm A}}}
\newcommand\thB{\ensuremath{\theta_{\rm B}}}
\newcommand\mfobs{\ensuremath{m_{\rm obs}}}
\newcommand\mfint{\ensuremath{m_{\rm int}}}
\newcommand\pgap{\ensuremath{p_{\rm gap}}}
\newcommand{\thetap}{\ensuremath{\theta_+}}
\newcommand\thetam{\ensuremath{\theta_-}}
\newcommand\tauI{\ensuremath{\tau^{\rm I}}}
\newcommand\tauS{\ensuremath{\tau^{\rm S}}}
\newcommand\phip{\ensuremath{\phi_{\rm p}}}
\newcommand\sigmam{\ensuremath{\sigma_{\rm m}}}
\newcommand\muth{\ensuremath{\mu_{\rm th}}}
\newcommand\mumin{\ensuremath{\mu_{\rm min}}}
\newcommand\siglim{\ensuremath{\sigma_{\rm lim}}}
\newcommand\zf{\ensuremath{z_{\rm f}}}
\newcommand\perGpcyr{\ensuremath{\mathrm{Gpc}^{-3}\,\mathrm{yr}^{-1}}}
\newcommand\tninety{\ensuremath{t_{90}}}
\newcommand{\be}{\begin{equation}}
\newcommand{\ee}{\end{equation}}
\newcommand{\ba}{\begin{eqnarray}}
\newcommand{\ea}{\end{eqnarray}}


\definecolor{orange}{rgb}{1,0.5,0}
\newcommand{\aipp}[1]{\textcolor{magenta}{[AIPP: #1]} }
\newcommand{\gps}[1]{\textcolor{orange}{[GPS: #1]} }

\begin{abstract}
Upcoming wide-field time-domain surveys, such as the Vera C. Rubin Observatory's Legacy Survey of Space and Time (LSST) are expected to discover up to two orders of magnitude more strongly lensed supernovae per year than have so far been observed. Of these, Type IIn supernovae have been predicted to be detected more frequently than any other supernova type, despite their small relative detection fraction amongst non-lensed supernovae. However, previous studies that predict a large population of lensed Type IIn supernova detections model their time evolving spectrum as a pure blackbody. In reality, there is a deficit in the UV flux of supernovae relative to the blackbody continuum due to line-blanketing from iron-group elements in the ejecta and scattering effects. In this work we quantify the effect of this UV suppression on the detection rates by LSST of a simulated population of strongly lensed Type IIn supernovae, relative to a pure blackbody model, using a mock LSST observing run. With a blackbody model, we predict to detect $\sim$70 lensed Type IIn supernova per year with LSST. By modelling a similar UV deficit to that seen in superluminous supernovae, we recover 60 - 80\% of the detections obtained using a pure blackbody model, of which $\sim$10 detections per year are sufficiently bright ($m_\textrm{i} < 22.5$ mag) and detected early enough (> 5 observations before lightcurve peak) to enable high-cadence spectroscopic follow up.
\end{abstract}



\section{Introduction}\label{sec:intro}

When a supernova (SN) detonates behind a foreground galaxy or cluster, it can be gravitationally lensed to form multiple images \citep{einstein_1936, zwicky_1937}. These images are magnified, sometimes by a large amount \citep{refsdal_1964a} allowing us to observe SNe further than they are usually detected. Moreover, the multiple images of gravitationally lensed SNe travel along different paths and therefore arrive at different times. These time delays can be used to provide a measure of the present-day expansion rate of the Universe, the Hubble constant, $H_0$, \citep{refsdal_1964b} that is independent of the cosmic microwave background (CMB) and local distance ladder measurements \citep{treu_2016, Birrer_2019, suyu_2024a}. If a lensed SN is identified before all of its images arrive, the appearance of further images can be anticipated, allowing for detailed analysis of the early-time, pre-peak features of the SN \citep{suwa_2017, goldstein_2018}. These unique properties allow lensed SNe to be powerful and distinct tools for understanding SN physics, stellar environments at high redshift and cosmology.

These events have so far been difficult to detect. To date, only eight lensed SNe with multiple images have been discovered. Of these, six were lensed by galaxy clusters: `SN Refsdal' \citep{kelly_2015}, `SN Requiem' \citep{rodney_2021}, `AT2022riv' (Kelly et al. 2022), `C22' \citep{chen_2022}, `SN H0pe' \citep{frye_2024}, and `SN Encore' \citep{pierel_2024}. The other two: `iPTF16geu' \citep{goobar_2017} and `SN Zwicky' \citep{goobar_2023} were lensed by a single galaxy-scale lens. While useful $H_0$ constraints have been obtained by some of these individual events \citep{Grillo_2018, Grillo_2020, Kelly_2023, Grillo_2024, pascale_2024}, in order to resolve issues with systematics, such as lens mass distribution models, and to conduct an investigation into the SN population at high redshift, a much larger sample of lensed SNe is needed. Fortunately, the next generation of telescopes, such as the Nancy Grace Roman Space Telescope \citep[e.g.][]{pierel_2021} and in particular the Vera C. Rubin Observatory \citep{Ivezic_2019}, promise to increase our discovery rate by a couple of orders-of-magnitude. The Rubin Observatory Legacy Survey of Space and Time (LSST) is expected to discover several hundred lensed SNe \citep{Oguri2010, goldstein_2019, wojtak_2019, arendse_2024, Sainz_de_Murieta_2024, Abe_2025}. 

Of these discoveries, the SN type that dominates is predicted to be Type IIn supernova (SNIIn). \citet{goldstein_2019} predict that LSST will observe 209.32 lensed Type IIn SNe per year, accounting for 55$\%$ of all lensed SN detections by LSST. \citet{wojtak_2019} similarly predict 210 lensed Type IIn SNe detections per year, accounting for 61$\%$ of their total lensed SN rate and 80$\%$ of their lensed core-collapse SN (CCSN) rate. This is in contrast with non-lensed Type IIn SNe being relatively uncommon, making up $\sim$ $5-10\%$ of non-lensed CCSNe \citep{Li_2011_II, Graur_2017, cold_2023}. 

Type IIn SNe are interesting astrophysical phenomena for a number of reasons. These SNe are part of the larger class of interacting SNe that are sometimes characterised using an 'n' in the name of their sub-class (e.g. IIn, Ibn, Icn), which refers to the narrow hydrogen emission lines in their spectrum. These narrow lines are indicative of a massive slow-moving circumstellar material (CSM) present around the exploding star \citep{Schlegel_1990, Schlegel_1996, Filippenko_1997}. This CSM is typically produced by non-terminal outbursts from the massive star prior to the SN. The conversion of kinetic energy into radiation from the shock between the SN ejecta and the CSM is the main source of the SN luminosity. This process is surprisingly efficient as Type IIn SNe are typically bright compared to Type Ia SNe and other CCSNe. Their luminosity covers a wide range, depending on the CSM mass available to decelerate the ejecta, from relatively faint SNe to the brightest superluminous SNe \citep{Nicholl_2020}.

The SN engine within the CSM is typically generated through core-collapse, but can be thermonuclear \citep{Silverman_2013, kool__2023}, or even a non-terminal outburst \citep{foley_2011, pastorello_2013}. If the CSM is sufficiently dense, the emission from the SN engine itself can be obscured \citep[e.g.][]{Smith_2008}. This makes it hard to classify these SNe in terms of their progenitors without better models of how different SN engines interact with different CSM structures, and what signatures of these interactions can be seen through observations. Producing a large catalogue of well sampled Type IIn SNe will allow us to develop and test models and gain a better picture of the diversity of sources and potential sub-classes \citep[e.g.][]{Ransome_2024, Hiramatsu_2024}. 

In addition, detecting Type IIn SNe can give us insights into the final stages of the lives of massive stars. The processes that occur before a star goes supernova, in particular mass-loss mechanisms, are difficult to characterise \citep{Smith_2014}. Mass loss rates previously calculated for SNe IIn range from above $10^{-3} \textrm{M}_{\odot} \,\textrm{yr}^{-1}$ \citep{Moriya_2014} to the staggering $2-7 \textrm{M}_{\odot} \, \textrm{yr}^{-1}$ in the case of SN ASASSN-14il \citep{Dukiya_2024}, with typical CSM masses of order 1$M_\odot$ \citep{Ransome_2024}. Such mass loss rates could be formed by a luminous blue variable (LBV) progenitor undergoing an eruptive mass-loss episode pre-SN \citep{Humphreys_1994, Ransome_2024}. Binary formation channels can also explain the mass-loss rates required to form a dense CSM, even in lower mass stars \citep{sana_2012, chevalier_2012}. 

The density and spatial structure of the CSM can also affect the shape of the SN light curve, the strength of emission lines, and can produce asymmetries such as double peaks \citep{Andrews_2017, Andrews_2018}. Aspherical CSM around Type IIn SNe is common, and can be constrained by their early-time ultraviolet (UV) properties \citep{Soumagnac_2020}. Many of these SNe only exhibit CSM-interaction features early in their evolution \citep[e.g.][]{Fassia_2001, Yaron_2017} or late in their evolution \citep[e.g.][]{Chugai_2006, Milisavljevic_2015,Benetti_2018}, providing insight into the time-scales between mass loss episodes and the SN itself.

While the sample size will be much lower \citep[see e.g.][]{Ransome_2024}, gravitational lensing offers some distinct advantages compared to non-lensed observations in the study of Type IIn SNe. Firstly, an early identification of a lensed Type IIn SN allows us to anticipate further images and dedicate high-cadence observations of these images to investigate the early time-evolving structure of the CSM. Additionally, with gravitational lensing allowing us to access high redshift detections, we can also begin to study whether certain CSM compositions and progenitor channels are favoured at different cosmic epochs. Additionally, as highlighted in \cite{Soumagnac_2020}, the source frame UV spectrum, which is redshifted into the LSST optical filter bands for a typical lensed Type IIn SN detection, is key in determining the bolometric luminosity of the supernova and to constrain CSM properties, particularly any asphericity in the CSM.

Furthermore, gravitational time delays from lensed SNe provide direct measurements of the cosmological distance that are independent of distance-ladder measurements. This time-delay distance is primarily sensitive to $H_0$, but also weakly dependant on other cosmological parameters. Due to the scarcity of lensed SNe to date, cosmological inferences from time-delay measurements have been predominantly utilised using measurements of lensed quasars \citep[e.g.][]{wong_2019, Birrer_2020, shajib_2020}. However, \cite{Kelly_2016} applied this technique successfully to a lensed SN, SN Refsdal, to produce robust measurements of $H_0 = 66.6_{-3.3}^{+4.1} \rm km \,s^{-1} Mpc^{-1}$, and \cite{Grillo_2024} use SN Refsdal to obtain measurements for the dark energy density parameter, $\Omega_\Lambda = 0.76_{-0.10}^{+0.15}$ and the dark energy equation of state parameter, $w = -0.92_{-0.21}^{+0.15}$. Of note, SN Refsdal is classified as a luminous blue SN 1987A-like Type II SN \citep{Kelly_2016}. These SNe exhibit similar properties and similar lightcurves to Type IIn SNe, highlighting their potential application as cosmological probes. With LSST forecasted to increase our sample of lensed SNe by at least an order of magnitude, samples of time-delay measurements from lensed SNe are expected to produce competitive measurements of cosmological parameters and of $H_0$ in as little as three years \citep{arendse_2024}.

The reason that Type IIn SNe dominate rate predictions of lensed SNe, despite being uncommon amongst non-lensed SNe, is that they are more suited to lensed detections compared to other SNe. This is due to three main factors: their colour, their prevalence as a function of redshift and their intrinsic brightness. Firstly, Type IIn SNe are hotter, and therefore ``bluer'', than other CCSNe and Type Ia SNe, with their peak flux typically in the near-UV. This means that at higher redshifts, $z \sim 2$, this UV peak corresponds to an observer frame spectral peak in the most sensitive optical bands of LSST. Other SNe are cooler and ``redder'', so at this redshift their spectral peak is shifted into the infrared, where LSST is less sensitive. Secondly, the occurrence rate of Type IIn SNe as a function of redshift is assumed to mirror the star formation rate due to the short lifetime of their most common progenitors. Unlike Type Ia SNe, which are more prevalent in the nearby universe, this makes Type IIn SNe prevalent at $z\sim2$, beyond the flux limit of current imaging surveys without the aid of magnification from lensing. While this is also true for other CCSNe, a typical Type IIn SN is much brighter than other CCSNe and so requires less lensing magnification to be detected. Putting this all together, Type IIn SNe are seemingly the ideal source for detection with the aid of lensing, and therefore are predicted to dominate LSST lensed SN detection rates.

To obtain a rate of lensed Type IIn SNe by LSST, both \citet{goldstein_2019} and \citet{wojtak_2019} use a template for their SNIIn lightcurve and spectral model from \citet{gilliland_1999}. This template consists of a time-evolving blackbody spectrum fitted to optical observational data of SN 1999el \citep{dicarlo_2002}. This pure blackbody model is suitably accurate for the optical spectrum of a SNIIn. However, it can be seen that the UV spectra of Type IIn SNe is far fainter than that predicted by blackbody representations \citep[see e.g.][]{Roming_2012}. This is because UV photons are absorbed by iron-group elements typically found in SN ejecta \citep{walker_2012}. The luminosity in this range is also diluted by UV photons being scattered to longer wavelengths by the fast expanding ejecta \citep{pauldrach_1996, Bufano_2009}. This effect has only been confirmed in CSM-interacting SNe a handful of times, especially at early times, as observations of the UV spectra are reliant on proximity, early discovery and classification. The two landmark examples are SN1987A \citep{arnett_1989, Harkness_1990}, and the recent SN 2023ixf, the best-observed SNe with CSM interaction features to date \citep[see][and references therein]{Bostroem_2024}. This effect is also seen in observations of super-luminous SNe (SLSNe), which exhibit significant absorption lines in the far-UV \citep[e.g.][]{yan_2018}.  We refer to this phenomenon as UV suppression. Since the source-frame UV flux coincides with the most sensitive LSST optical bands for a typical lensed Type SNIIn, by accounting for this UV suppression, the overall predicted rates of lensed Type SNIIn detections will decrease.


In this work, we explore the effect of implementing a simple model to explore the impact of UV suppression on the overall detection rates of gravitationally lensed Type IIn supernovae by LSST. Throughout this work, we assume
a standard flat Lambda cold dark matter ($\lambda$CDM) model with
$H_0$ = 67.8 km $\textrm{s}^{-1} \textrm{Mpc}^{-1}$ and $\Omega_{\textrm{m}}$ = 0.308 \citep{planck18}.

We outline the model used to simulate our population of strongly lensed Type IIn supenovae in Section \ref{sec:Simulating}, our results are presented in Section \ref{sec:results} and our discussion and conclusions are found in Section \ref{sec:discussion} and Section \ref{sec:conclusion}.


\section[Simulating Lensed Type IIn SNe]{Simulating Lensed Type II\lowercase{n} SN\lowercase{e}}\label{sec:Simulating}

In this section, we describe the model used to
forecast the rates and properties of gravitationally lensed SNIIn detected by LSST. This framework can be adapted to other lensed astronomical transients and therefore for clarity we distinguish which components are specific to our investigation into SNIIn rates.

\subsection{The formalism}\label{sec:formalism}

The number of detections of a lensed transient per year can be expressed as:


\begin{equation}
  R_{\textrm{det}} = \int_{0}^{\infty} \textrm{d}\Lambda
  \int_{\mu_{\textrm{min}\left(z\right)}}^{\infty}\!\!\!\!\!\textrm{d}\mu
  \int_{z_{\textrm{min}}}^{z_{\textrm{max}}} \textrm{d}z \,\,\, 
  N_{\textrm{phys}}\!\left(z\right) \, \frac{\textrm{d}\tau}{\textrm{d}\mu dz} \,
  \phi\!\left(z, \Lambda \right).
  \label{eqn:Rdet}
\end{equation}
Here, $\mu_{\textrm{min}(z)}$ is the minimum gravitational magnification required to produce a detectable
signal from a source, and $\left(z_{\rm min}, z_{\rm max}\right)$ denotes the redshift range over
which the chosen telescope is sensitive to the source. $N_{\rm phys}\!\left(z\right)$ represents the total yearly rate of the transient event that occur in a redshift interval ${\textrm{d}z}$. This depends on the redshift evolution of the transient and is described in Section \ref{sec:redshift}. The second term, $\textrm{d}\tau / \textrm{d}\mu \textrm{d}z$ represents the differential source plane optical depth for each redshift interval, $\textrm{d}z$ and lens magnification interval $\textrm{d}\mu$. This term is described in Section \ref{sec:optical_depth}. The final term, $\phi\left(z\right)$, represents the fraction of all lensed sources of the chosen transient in a redshift shell $\textrm{d}z$ that are detected by a given survey. The parameter $\Lambda$ encapsulates the information about the source, instrument and observing conditions that renders the source more or less detectable, including (but not limited to) the source redshift, luminosity, lensing magnification, evolution over time, spectral properties, weather, moon phase, atmospheric effects, instrumental cadence and instrument sensitivity. The model used to obtain $\phi\left(z\right)$ for the population of lensed SNIIn detected by LSST is laid out in section \ref{sec:detection rates}.

\subsection{Redshift distribution}\label{sec:redshift}

The first term in equation \ref{eqn:Rdet} can be further broken down as:
\begin{equation}
  N_{\rm phys}\left(z\right) = R_0 \, g(z) \, \frac{\textrm{d}V}{\textrm{d}z} \cdot (1+z)^{-1}
  \label{eqn:Nj}
\end{equation}
where $R_0$ is the local comoving rate density of the source type, $g(z)$ is a function that describes how this rate density evolves with redshift and $\textrm{d}V$ is the comoving volume element corresponding to the redshift interval $\textrm{d}z$. The final term accounts for time dilation of the rates into the observer frame.

The local comoving rate density of SNIIn, $R_0=4.77 \cdot 10^{-6} \,\rm yr^{-1} \,\rm Mpc^{-3}$ \citep{cold_2023}. We assume that the rate density of SNIIn mirrors the star formation rate density due to the relatively short lifetime of their progenitors and therefore this local density evolves according to a double power law model:
\begin{equation}
  g(z)=C(1+z)^\alpha\left[1+\left(\frac{1+z}{1+\zpivot}\right)^{\alpha+\beta}\right]^{-1}
  \label{eqn:madau}
\end{equation}
where $\zpivot$ is the redshift at which the redshift evolution transitions from its low redshift regime of $g\propto(1+z)^\alpha$ to its high redshift regime of $g\propto(1+z)^{-\beta}$, and $C=1+(1+\zpivot)^{-(\alpha+\beta)}$ ensures that $g(z=0)=1$, following \cite{Callister_2020}. We adopt parameter values of $\alpha=2.7$, $\beta=2.9$, and $\zpivot=1.9$ which represent the evolution of the star formation rate density of the universe \citep{madau_2014}.

\subsection{Lensed supernovae}\label{sec:optical_depth}

Phrasing the optical depth to gravitational lensing in terms of magnification in Equation~\ref{eqn:Rdet} captures the magnifying power of the ensemble of galaxy, group, and cluster-scale lenses in the observable universe independent of their mass and structure \citep{Smith_2023}. This term has been shown via several methods to be well described by 
\begin{equation}
  \frac{\textrm{d}\tau}{\textrm{d}\mu \textrm{d}z}=\left(\frac{D}{31\,\rm Gpc}\right)^3\mu^{-3},
  \label{eqn:dtaudmu}
\end{equation}
where $D$ is the comoving distance to the redshift $z$ \citep[e.g.][]{haris_2018, robertson_2020}. When considering just the effects of gravitational magnification, a source of given luminosity at given redshift is detectable if it is magnified by at least the minimum required to render it detectable, $\mu_{\rm min}$. Therefore, it is informative to marginalise over $\mu$ to obtain:
\begin{equation}
  \frac{\textrm{d} \tau}{\textrm{d}z}\bigg{|}_{\mu>\mu_{\rm min}} = \int_{\mu_{\rm min}}^{\infty} \, \frac{\textrm{d}\tau}{\textrm{d}\mu \textrm{d}z} \,
  \textrm{d}\mu \ =
  \frac{1}{2} \left(\frac{D}{31\,\rm Gpc}\right)^3 \mu_{\rm min}^{-2}.
  \label{eqn:tau}
\end{equation}
This equation describes the fraction of the celestial sphere at a comoving distance $D$ that is magnified by $\mu > \mu_{\rm min}$.


\begin{figure}
    \centering
    \hspace{-0.5cm}
    \includegraphics[width=1.05\columnwidth]{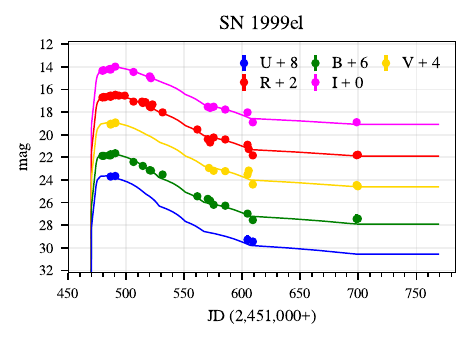}
    
    \caption{Lightcurve fits for SN 1999el used to produce the blackbody Type IIn SN template.}
    \label{fig:lightcurve}
\end{figure}


The signals from gravitationally lensed SNe detectable in the focal plane of modern telescopes are also modified by the temporal and angular separations of pairs lensed images. Specifically, if the signals from multiple images overlap in time and position, then the photometric measurements of the lensed SN can be brighter and extend over a longer period of time. Temporal and angular separation depend on the mass and internal structure of the lens and the nature of the lensing catastrophe responsible for the multiple images. For simplicity, in our analysis we only consider image pairs. We use the Singular Isothermal Sphere (SIS) model for galaxy-scale lenses, which produces only double-image systems. For galaxy-, group- and cluster-scale lenses, where image quads can form near a fold caustic, we approximate them by considering only the two brightest images, as they dominate the observed signal. Following \cite{Smith_2023}, we use the image separation and arrival time difference relations for double images formed by galaxy scale lenses and fold image pairs formed by galaxy/group/cluster-scale lenses. For each lens archetype, we assume a fixed Einstein radius, $\theta_{\rm E}$, and a fixed value for the slope, $\eta_{\rm E}$, and density, $\kappa_{\rm E}$ of the lens at the Einstein radius, as listed in Table~\ref{tab:lens_parameters}. This simplification is justified by the fact that the distribution of Einstein radii and the slope and density, is implicitly accounted for in the optical depth distribution from \citet{robertson_2020}. As a result, variations in these parameters primarily influence time delays and image separations, and are expected to have a minimal impact on the final detection rate outcomes.
Moreover, \citeauthor{Smith_2023} showed that the relations and parameter values we adopt for image separations and time delays capture the behaviour of time domain lenses detected thus far. To describe galaxy-scale lenses that form two images, we use
\begin{equation}
  \Delta\theta = 2\theta_{\rm E},
  \label{eqn:image_separation_SIS}
\end{equation}
\begin{equation}
  \Delta t = 92\textrm{d} \bigg[\frac{\theta_{\rm E}}{1''}\bigg]^2\bigg[\frac{\mu_{\rm p}}{4}\bigg]^{-1} \bigg[\frac{D_{\Delta t}}{3.3 \textrm{Gpc}}\bigg],
  \label{eqn:time_delay_SIS}
\end{equation}
where $\theta_{\rm E}$ is the Einstein radius of the lens, $\mu_{\rm p} = |\mu_1| + |\mu_2|$ is the sum of the absolute value of the magnification of each image in the pair, $D_{\rm \Delta t}=D_{\rm L}D_{\rm S}/D_{\rm LS}$, $D_{\rm S}$ and $D_{\rm L}$ are the comoving distances to the source and lens respectively, $D_{\rm LS}=D_{\rm S}-D_{\rm L}$, and $\textrm{D}_{\Delta t}=3.3\,\rm Gpc$ corresponds to $\zl=0.5$ and $\zs=1.6$.

To capture the fold caustic contributions to quad images formed by galaxy-scale lenses and the fold-image pairs that dominate group/cluster-scale lenses we instead use
\begin{equation}
  \Delta\theta = \frac{\theta_{\rm E}}{\mu\,\eta_{\rm E}\,\kappa_{\rm E}(1-\kappa_{\rm E})},
  \label{eqn:image_separation_fold}
\end{equation}
\begin{equation}
  \Delta t = 3.9\textrm{d} \bigg[\frac{\theta_{\rm E}}{1''}\bigg]^2 \bigg[\frac{\mu}{4}\bigg]^{-3} \bigg[\frac{\eta_{\rm E}}{1}\bigg]^{-2} \bigg[\frac{\kappa_{\rm E}}{0.5}\bigg]^{-2} \bigg[\frac{D_{\Delta t}}{3.3 \textrm{Gpc}}\bigg]
  \label{eqn:time_delay_fold}
\end{equation}
where $\eta_{\rm E}$ and $\kappa_{\rm E}$ are the slope and density of the lens at the mid-point of the fold image pairs and $\mu$ is the single-image magnification (i.e. $\mu = \mu_1 = \mu_2$). For derivations of these relationships, see Appendix~A of \citet{Smith_2023} and references therein. The representative values of $\theta_{\rm E}$, $\kappa_{\rm E}$ and $\eta_{\rm E}$ that we use for galaxy, group and cluster-scale lenses are summarised in Table \ref{tab:lens_parameters}.


\begin{figure}
    \centering
    \hspace{-0.5cm}
    \includegraphics[width=1.05\columnwidth]{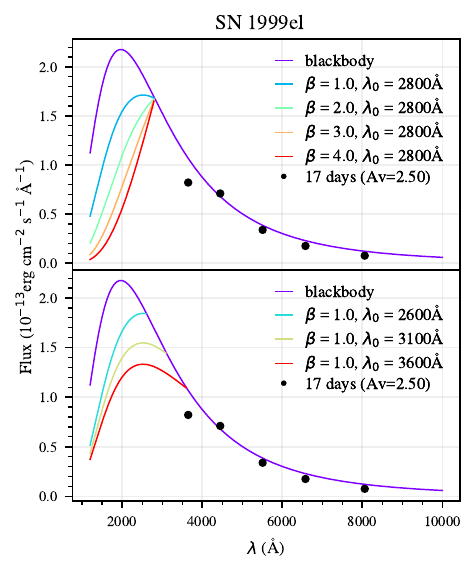} 
    \caption{Spectral fits for SN 1999el for a range of UV suppression models. Top panel shows how the spectrum changes for different $\beta$ values, bottom plane shows how the spectrum changes for different $\lambda_0$ values. Black dots represent flux obtained 17 days after the supernova assuming $A_{\rm V}$ = 2.50 mag.}
    \label{fig:spectra}
\end{figure}

\subsection{Type IIn Supernova Lightcurves and UV Spectra}\label{sec:lightcurves}

To generate a rest-frame lightcurve, we use \textsc{SNCosmo} \citep{barbary_2024} and its built-in IIn lightcurve model \textsc{Nugent-SN2n} \citep{gilliland_1999}. This template is based on data of SN1999el \citep{dicarlo_2002} and is the same model used for lensed SNIIn detection rates in \cite{goldstein_2019} and \cite{wojtak_2019}. A more detailed justification for our choice of lightcurve model is provided in Section~\ref{subsec:heterogeneity}. Throughout this paper, this template is referred to as the blackbody model. The lightcurve used to create this model is shown in Figure \ref{fig:lightcurve}.

To produce models that incorporate UV suppression effects, we modify this blackbody with a two-parameter power law model: 

\begin{equation}
  f_{\lambda} = 
  \begin{dcases}
    f_{\lambda, \mathrm{BB}}\left(\frac{\lambda}{\lambda_0}\right)^{\beta}, & \text{if } \lambda < \lambda_0 \\
    f_{\lambda, \mathrm{BB}}\vphantom{\left(\frac{\lambda}{\lambda_0}\right)^{\beta}}, & \text{else }
  \end{dcases}
  \label{eqn:UV_suppression}
\end{equation}

where $f_{\lambda, BB}$ is the flux from the blackbody model, $\lambda_0$ is the wavelength below which the suppression occurs, and $\beta$ is the power law index governing the strength of the suppression. This model was introduced by \citet{yan_2018} and itself builds upon a model from \citet{nicholl_2017}. \citeauthor{yan_2018} use this model to fit the deficit in the UV continuum of SLSNe. This model is simple and adaptable, and is based on observed properties of the continuum rather than analytic models. This simplicity is suitable for the analysis in this paper, as it allows us to characterise the strength and location of the suppression. We generate templates spanning $0.5 \leq \beta \leq 4.0$ and $2600 \leq \lambda_0 \leq 3600$. These ranges are wide and extend beyond the values obtained by \cite{yan_2018} of $0.6 \leq \beta \leq 3.0$ and $2800 \leq \lambda_0 \leq 3500$ for SLSNe. Since the sample size in \citeauthor{yan_2018} is small, and applies to a different class of SN, we use a larger range to account for the possibility of decreased or increased suppression. Some example spectra produced by this UV model for a range of UV suppression parameters, overlaid on the blackbody fit for SN1999el, are shown in Figure \ref{fig:spectra}.

\begin{table}
	\centering
	\caption{Parameter values for different lens archetypes used to calculate the arrival time differences and angular separations of the gravitationally lensed supernova images.}
	\label{tab:lens_parameters}
	\begin{tabular}{lcccc} 
		\hline
		Lens & $\theta_{\rm E}$ & $\kappa_{\rm E}$ & $\eta_{\rm E}$ \\
		\hline
        Galaxy-scale & $1''$ & $0.5$ & $1$\\
        Group-scale & $3''$ & $0.65$ & $0.75$\\
        Cluster-scale & $10''$ & $0.8$ & $0.5$\\
		\hline
	\end{tabular}
\end{table}

\subsection{The Vera C. Rubin Observatory}\label{sec:rubin}

The Vera C. Rubin Observatory is an upcoming survey facility located on Cerro Pach\'on in Chile. It will conduct the Legacy Survey of Space and Time (LSST), taking multiband \textit{ugrizy} images of $\sim 20 \, 000 \, \rm deg^2$ of the sky to a limiting magnitude of $\sim$ 24. Observations will be taken using the Simonyi Survey Telescope. The survey is planned to run for 10 years commencing in late 2025. Due to its impressive depth and rapid sky coverage, LSST is the most promising upcoming survey for detecting lensed transient events, as highlighted by the forecasted discovery rates of several hundred lensed transients a year \citep{Oguri2010, goldstein_2019, wojtak_2019, arendse_2024}.

LSST will operate with several survey modes. There are two main survey modes that we consider: the Wide-Fast-Deep (WFD) survey which accounts for $\sim\rm90 \%$ of observing time and the deep drilling fields (DDFs). These are the two modes that take up the bulk of the survey time, are nominal for lensed SN detections, and are not located in areas of high dust extinction such as the galactic plane. The WFD survey is currently set to use a rolling cadence, where certain areas of the WFD footprint will be assigned more visits, with this area of focus 'rolling' across the sky over time. This will improve the sampling density and therefore quality of fitted light curves for transients in the high-cadence, active, regions. The rolling cadence begins roughly 1.5 years into the survey to allow the first season of observations to be uniform. \citet{alves_2023} show that the active region provides a $\rm25 \%$ improvement in the classification of unlensed SNe types relative to the background region. However, a drawback to the rolling cadence is that there is a decreased likelihood of detecting events in the background regions. This is particularly impactful for rare, and fast, transients. \citet{arendse_2024} show that, despite this, there is only a small decrease to the number of lensed SNIa per year when using a rolling cadence. While we do not investigate the effect of a rolling vs non-rolling cadence in this work, we expect a similar small decrease for rates of lensed SNIIn detections.

Our analysis utilises the v3.4 observing strategy. v3.4 should be comparable to previous versions from v3.0 onwards, with the biggest change occurring in v3.3 with an updated mirror coating that sacrifices \textit{u} band sensitivity to boost sensitivity in the \textit{grizy} bands. While lensed Type IIn SNe are blue, they are most commonly found and detected at a redshift where the peak wavelength is found in the redder bands, and so this change should result in a slightly increased number of lensed SNIIn detections.

\begin{table}
    \begin{threeparttable}
	\centering
	\caption{Parameter distributions for lensed SNIIn. $\mathcal{N}\left(\mu, \sigma\right)$ represents a normal distribution with mean $\mu$ and standard deviation $\sigma$, and $\mathcal{U}\left(x, y\right)$ represents a uniform distribution from $x$ to $y$.}
	\label{tab:parameters}
	\begin{tabular}{lc} 
		\hline
		Parameter & Distribution\\
		\hline
		Redshift & $z \sim \mathcal{U}\left(0, 6\right)$ \tnote{a}\\
		Magnification & $\mu \sim \mathcal{U}\left(2, 50\right)$ \tnote{b}\\
            Magnitude & $M_{\rm B} \sim \mathcal{N}\left(-19.05, 0.5\right)$ \tnote{c}\\
            Right Ascension (rad) & $\alpha \sim \mathcal{U}\left(-\pi, \pi\right)$\\
            Declination (rad) & $\delta = \rm arcsin (\theta), \theta \sim \mathcal{U}\left(-1, 0.64\right)$ \tnote{d}\\
		\hline
	\end{tabular}
        \begin{tablenotes}
            \footnotesize
            \item[a] The sample redshift distribution is uniform to accurately sample the tails of the physical redshift distribution. The true physical redshift distribution is contained within the $N_{\rm phys}$ term in equation \ref{eqn:Rdet}, as explained in section \ref{sec:redshift}.
            \item[b] The sample magnification distribution is uniform to accurately sample the tails of the physical magnification distribution. The true physical magnification distribution is contained within the $\frac{d\tau}{d\mu dz}$ term in equation \ref{eqn:Rdet}, as explained in section \ref{sec:optical_depth}. An upper limit of 50 is used as sources with magnifications above 50 are improbable and so contribute little to the overall detection rates.
            \item[c] This distribution is the same as that used in \citet{goldstein_2019}.
            \item[d] arcsin$(0.64) = 40\deg$, which is approximately the maximum declination observed by LSST. 
      \end{tablenotes}
    \end{threeparttable}  
\end{table}

\subsection{Calculating Detection Rates}\label{sec:detection rates}


To calculate the annual number of lensed Type SNIIn detections and their properties, we start by generating a sample of 200 000 unique lensed Type IIn SNe. Each SN is assigned a value from each of the distributions in Table \ref{tab:parameters}. For the sake of comparison, we use the same distribution for the peak rest-frame B-band absolute magnitude \citet{Li_2011_II} as used in \citet{goldstein_2019}. The template lightcurves are scaled up uniformly to a peak rest-frame magnitude $M_{\textrm{B}}$, magnified by $\mu$, and k-corrected to produce observer frame lightcurves for the source at a redshift $z$ and sky-position of $\left(\alpha, \beta\right)$. 

The next step involves determining whether each lensed SN is detected by LSST. Firstly, if the sky position of the SN falls outside the LSST footprint then the SN is marked as undetected. Following \citet{arendse_2024}, in order to distinguish between the galactic plane region and the WFD and DDF fields, we use a threshold number of sky visits over 10 years of LSST observations. Regions with $N_{\rm visits} < \rm 400$ are part of the galactic plane and polar regions, and any observations that occur in these regions are discarded. 


\begin{figure*}
    \centering
        \begin{subfigure}[t]{0.49\textwidth}
        \centering
        \hspace{-0.9cm}
        \includegraphics[width=\linewidth]{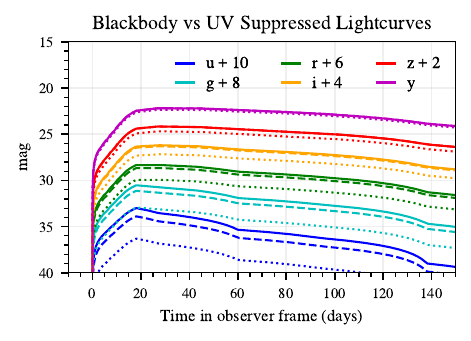} 
    \end{subfigure}
    \hfill
    \begin{subfigure}[t]{0.49\textwidth}
        \centering
        \hspace{-0.9cm}
        \includegraphics[width=\linewidth]{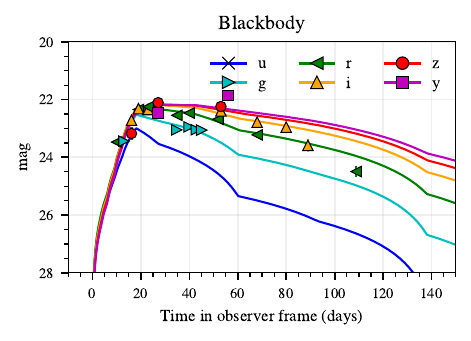} 
    \end{subfigure} \\
    \begin{subfigure}[t]{0.49\textwidth}
        \centering
        \hspace{-0.9cm}
        \includegraphics[width=\linewidth]{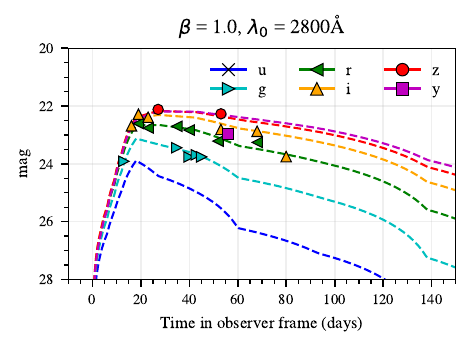} 
    \end{subfigure}
        \hfill
    \begin{subfigure}[t]{0.49\textwidth}
        \centering
        \hspace{-0.8cm}
        \includegraphics[width=\linewidth]{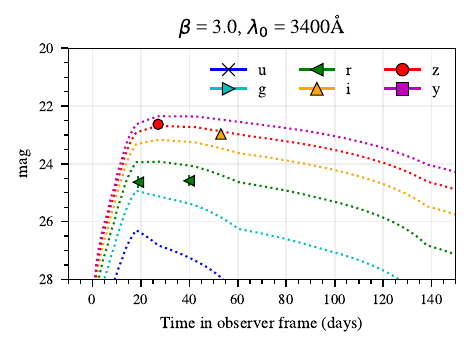} 
    \end{subfigure}
    \caption{A supernova in our sample at $z = 2$ with $M_{\textrm{B}} = -19.05$ and $\mu = 10$. Top left: Lightcurves for a blackbody model (solid), a low suppression $\beta = 1.0, \lambda_0 = 2800$Å model (dashed) and a high suppression $\beta = 3.0, \lambda_0 = 3400$Å model (dotted). From top-right going clockwise: the lightcurves for each of these models with markers corresponding to LSST observations across different filter bands.}
    \label{fig:BBvsUV}
\end{figure*}


Each SN is also assigned a random start date, between the first day of observations to the end of year 3. To find the set of times at which SNe at different locations are observed, along with the relevant information about each observation, we use the Rubin Operations Simulator (\textsc{OpSim}), which simulates a mock LSST observing run over the 10-yr duration of the survey \citep{Delgado_2014, Delgado_2016, Naghib_2019}. Detailed metadata about each simulated pointing of the telescope is then accessed using \textsc{OpSimSummary} \citep{Biswas_2020}. From the \textsc{OpSim} database, we obtain the observing times and associated filters, the mean $5\sigma$ depth ($m_5$), and point-spread functions (psf) of each observation of each SN in our sample. Using the $5\sigma$ depth, we compute the $1\sigma$ noise on the SN flux:
\begin{equation}
  \sigma = \frac{10^{\, 0.4\left(m_0 - m_5\right)}}{5}
  \label{eqn:sigma}
\end{equation}
where $m_0$ is the instrument zero-point. This $1\sigma$ noise encapsulates contributions from the sky brightness, airmass, atmospheric effects, and the psf. The simulated lightcurve flux, $f$, evaluated at each of the observation times is then perturbed using gaussian noise with a width equal to $\sigma$. The error on the observed magnitude is then:  

\begin{equation}
  \sigma_m = \left| \frac{-2.5\sigma}{f\ln{10}} \right|.
  \label{eqn:sigma_M}
\end{equation}

For each simulated SN and each UV suppression model we now have a sequence of observations, with an apparent magnitude and the filter the observation was taken in. We consider a lensed IIn to be detected if it is observed at least 5 times across any of the LSST filters. These observations must have a signal-to-noise ratio greater than 5 and can be across one image, both images, or a single unresolved image. Figure \ref{fig:BBvsUV} illustrates a source that meets this detection criteria for a blackbody model and a low UV suppression model ($\beta = 1.0, \lambda_0 = 2800$Å), but falls short with only 4 observations for a high suppression model ($\beta = 3.0, \lambda_0 = 3400$Å). This criterion of detection should be interpreted as an upper limit to the number of lensed Type IIn SNe we will \textit{identify} with LSST. It allows us to confidently conclude that a transient was observed, but for many of our detections we will not have the information required to classify them as a lensed Type IIn SN. In Section \ref{sec:observation_quality} we investigate how the total number of detections changes as the number of observations required to constitute a detection changes. The more observations an event has, the more likely we are to be able to identify the transient as a lensed SNe IIn and the more we expect to learn about the event.


\begin{figure*}
    \textbf{\large Overall Yearly Detection Rates}\par\medskip
    \centering
        \begin{subfigure}[t]{0.49\textwidth}
        \centering
        \includegraphics[width=\linewidth]{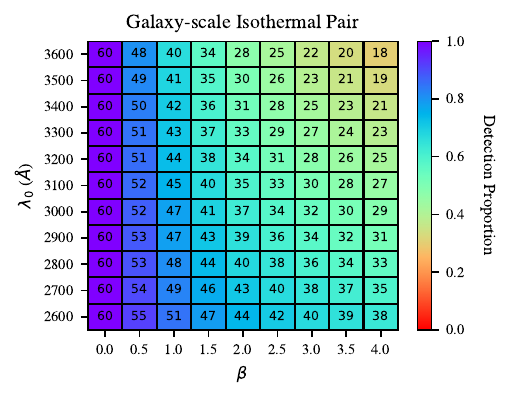} 
    \end{subfigure}
    \hfill
    \begin{subfigure}[t]{0.49\textwidth}
        \centering
        \includegraphics[width=\linewidth]{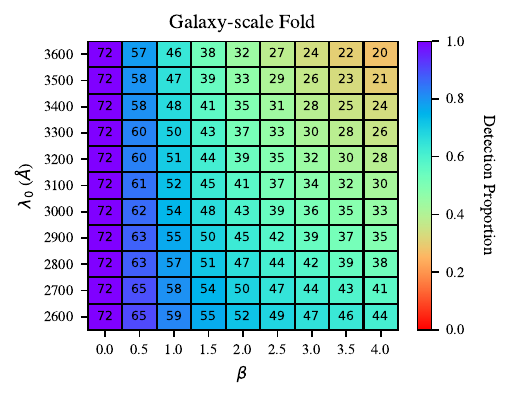} 
    \end{subfigure} \\
    \begin{subfigure}[t]{0.49\textwidth}
        \centering
        \includegraphics[width=\linewidth]{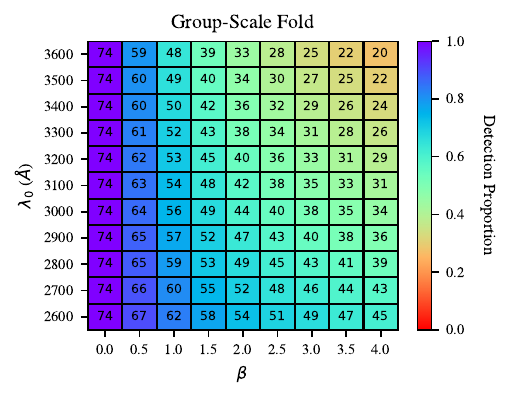} 
    \end{subfigure}
        \hfill
    \begin{subfigure}[t]{0.49\textwidth}
        \centering
        \includegraphics[width=\linewidth]{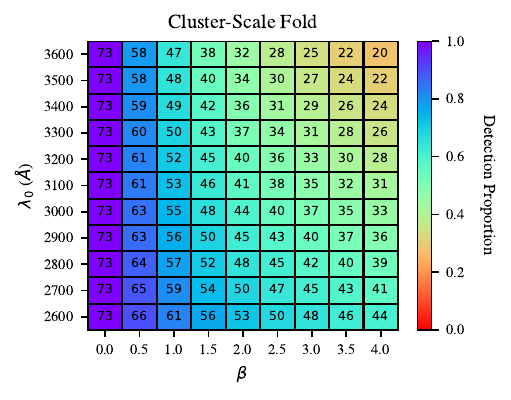} 
    \end{subfigure}
    \caption{The total yearly LSST detection rate of lensed Type IIn SNe for a range of UV suppression models. Each square in the grid corresponds to a different combination of power law index, $\beta$, and cutoff wavelength, $\lambda_{0}$. The left hand column corresponds to $\beta = 0$, i.e. a pure blackbody model. The colour bar represents the fraction of detections for each model relative to the detections obtained using a pure blackbody model for the given lens model. The four panels each show results for a different lens archetype. From the top-left going clockwise, the lens models represent images formed by: galaxy-scale isothermal lenses, galaxy-scale fold caustics, cluster-scale fold caustics and group-scale fold caustics. These models are described in Section \ref{sec:optical_depth}.}
    \label{fig:detection_heatmaps}
\end{figure*}


\subsection{Gold Sample for Follow-up Observations}\label{sec:gold sample}

In order to obtain study Type IIn SNe in detail, it is important to identify them early in their evolution. This is primarily to be able to schedule dedicated, high-cadence spectroscopic follow-up. The benefits of this are two-fold: a) high-cadence spectroscopy, particularly around the peak, is vital to investigate the time-varying characteristics of Type IIn SNe and their CSM properties and b) spectroscopy is required to determine the SN type and redshift, which can also help to constrain whether a SN is magnified, and therefore, lensed. Furthermore, confirmed lensed Type IIn SNe can be used to obtain time-delay measurements if future images are observed. We borrow the definition of a ``gold sample'' from \cite{arendse_2024}, that meet the following criteria:

\begin{enumerate}[labelwidth=*, labelsep=0.3em, leftmargin=0.3in, rightmargin=0.3in]
    \item $N_{\rm prepeak} > 5$ in at least 2 filters,
    \item $m_{\rm i} < 22.5$,
    \item $\Delta t > 10$ days,
\end{enumerate}
with $N_{\rm prepeak}$ is the number of detections with signal-to-noise ratio > 3 before the SN peak and $m_{\rm i}$ the apparent i-band magnitude at peak.

The first criterion allows for spectroscopic follow-up when the
lensed SN is close to its brightest, and provides time
for scheduling high-resolution follow-up. The second criterion ensures that the lensed SN is bright enough to get a classification spectrum with a 4-m class telescope, e.g. 4-metre MultiObject Spectroscopic Telescope \citep[4MOST; ][]{4MOST} with
exposure times of 1 hour. Alternatively, with shorter exposure times, the spectra can be
obtained with instruments on 8m class telescopes, e.g. the Gemini
Multi-Object Spectrographs \citep[GMOS; ][]{Gemini}. Applying cuts based on these two criteria to our population of detections produces our ``spectroscopic gold sample'' of SNe for which we can feasibly investigate their time-evolving CSM properties or more general spectroscopic properties. The third criterion, applied alongside the first two, allows for ample time to classify the supernova, determine that it is lensed, and predict the emergence time of a future image before the image appears. This allows high cadence follow up of the original and future image to obtain time-delay measurements. SNe in our sample that meet all three criteria form our ``time-delay gold sample''. These are SNe for which we can potentially obtain accurate time delay measurements that can be used to infer the value of $H_0$ and other cosmological parameters.

\section{Results}\label{sec:results}

\subsection{Total Detections}\label{sec:total_detections}

Following the procedure described in Section \ref{sec:detection rates}, we obtain total yearly detections of strongly lensed Type IIn SNe for a range of different UV suppression models. These are summarised in figure \ref{fig:detection_heatmaps}. We find that $\sim$ 70 lensed Type IIn SNe are detected per year by LSST for a pure blackbody model. This is a third of the total number predicted by \cite{goldstein_2019} and \cite{wojtak_2019}. However, unlike these studies, we are considering the full LSST v3.4 observing strategy, which is comparable to the v3.0 strategy used by \cite{arendse_2024} (see Section \ref{sec:rubin}). \citeauthor{arendse_2024} find that, for lensed Type Ia SNe, when the LSST observing strategy is implemented, only 46\% of doubles are retained from the total detections calculated using the methods used in \citet{wojtak_2019}. For simplicity, in our analysis, we are only considering image pairs, and therefore it is reasonable to expect a similar decrease in total detections for lensed Type IIn SNe. In addition, our choice of five observations being required for a source to be classed as detected is stricter than the detection criteria used in \cite{arendse_2024}, leading to a further decrease in rates which is reflected in our values. As such, our results can be interpreted as conservative estimates. The effects of relaxing these detection criteria is explored in Section \ref{sec:observation_quality}.

With regards to the effects of UV suppression, it can be seen that even for the lowest suppression model considered, $\beta = 0.5$ and $\lambda_0 = 2600$\r{A}, there is a $\sim$10\% drop in the detection rates. The detection rates decrease smoothly as $\beta$ and $\lambda_0$ increase, with the highest suppression model of $\beta = 4.0$ and $\lambda_0 = 3600$\r{A} only recovering  $\sim$30\% of the detections made using a pure blackbody model. Based on the suppression due to line blanketing seen in SLSNe, a typical SLSN UV suppression model for fixed $\lambda_0 = 2800$\r{A} has $\beta$ ranging from 0.8 to 3.0 \citep{yan_2018}. This spans a range where roughly 80\% to 60\% of blackbody detections are recovered. It is unclear how much the interaction with a dense CSM may increase or decrease the amount of UV suppression for a SNIIn compared to a SLSN. However, even if the suppression is more powerful, it does not impact the detection rates enough to significantly harm the prospects of obtaining a useful sample of lensed Type IIn SNe with LSST. Even in the unlikely scenario where the suppression is extreme, the population is not extinguished, with $\sim$20 detections per year. This is still an order of magnitude more than the total lensed SN detections to date.

For the isothermal image-pair model, we obtain fewer detections across all models. This is due to the fact that, for an image pair produced by an isothermal lens, each image has a different value of magnification. We prescribe the drawn value of $\mu$ to be the image with the greater magnification, $\mu_1$, and therefore the second image has a magnification of $\mu_2 = \mu - 2$. For the image pairs produced by fold-caustic models, both images in a pair have the same magnification equal to $\mu$. Despite this, the overall trend observed in the proportional decrease in detections for the UV suppression models relative to the blackbody model is the same.


\begin{figure*}
    \centering
    \begin{subfigure}[t]{0.49\textwidth}
        \centering
        \hspace{-0.8cm}
        \includegraphics[width=\linewidth]{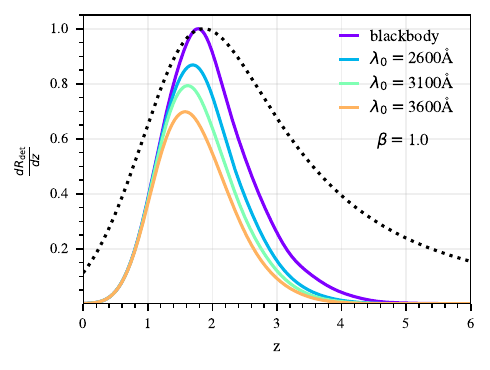} 
    \end{subfigure}
    \begin{subfigure}[t]{0.49\textwidth}
        \centering
        \hspace{-1.2cm}
        \includegraphics[width=\linewidth]{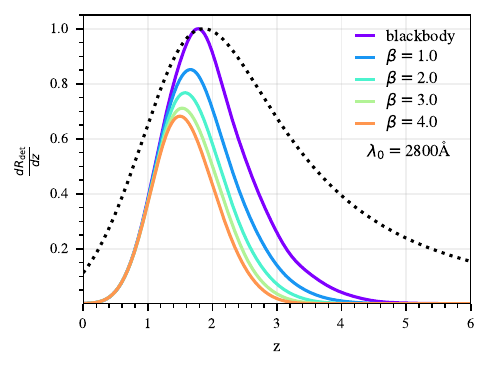} 
    \end{subfigure}
    \caption{The distribution of detected sources as a function of redshift, $z$, for a range of different UV suppression models using a galaxy-fold lens model. The detections are normalised so that a value of 1 corresponds to the peak number density of detections for a pure blackbody model. The dotted curve represents the evolution of the cosmic Type IIn SN rate density. Left: The effect of changing $\lambda_{0}$ for a fixed $\beta = 1.0$. Right: The effect of changing $\beta$ for a fixed $\lambda_{0} = 2800$\r{A}.}
    \label{fig: galaxy_redshift_histogram}
\end{figure*}


\begin{figure*}
    \centering
    \begin{subfigure}[t]{0.49\textwidth}
        \centering
        \hspace{-0.8cm}
        \includegraphics[width=\linewidth]{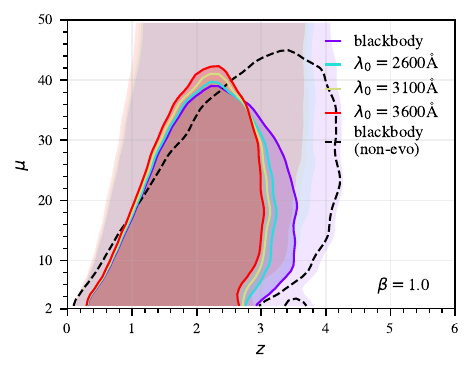} 
    \end{subfigure}
    \hfill
    \begin{subfigure}[t]{0.49\textwidth}
        \centering
        \hspace{-1cm}
        \includegraphics[width=\linewidth]{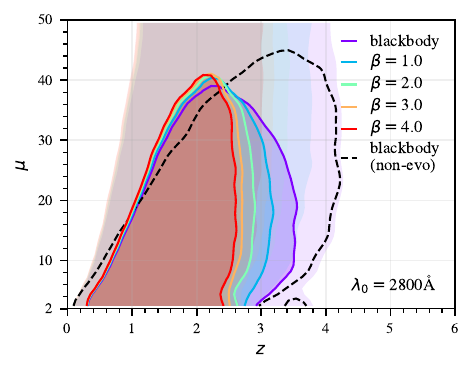} 
    \end{subfigure}
    \caption{The distribution of detected sources as a joint function of redshift, $z$, and magnification, $\mu$, for a range of different UV suppression models using a galaxy-fold lens model. Contours encloses 90\% of the predicted lensed detections, and the shaded areas extend
    to 99\% to visualise the tails of the respective distributions. The dashed curve represents 90\% of the distribution of detections for a non-evolving source population using the blackbody model. Left: The effect of changing $\lambda_{0}$ for a fixed $\beta = 1.0$. Right: The effect of changing $\beta$ for a fixed $\lambda_{0} = 2800$\r{A}.}
    \label{fig: redshift_magnification}
\end{figure*}


\begin{figure*}
    \centering
    \begin{subfigure}[t]{0.49\textwidth}
        \centering
        \hspace{-0.8cm}
        \includegraphics[width=\linewidth]{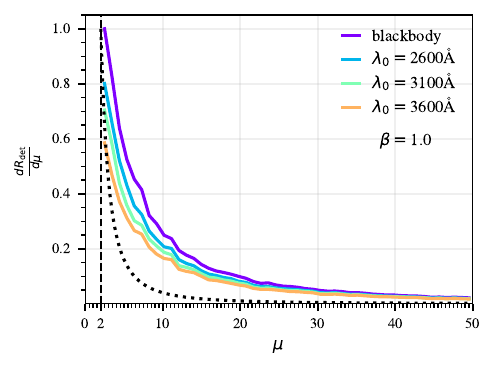} 
    \end{subfigure}
    \begin{subfigure}[t]{0.49\textwidth}
        \centering
        \hspace{-1.2cm}
        \includegraphics[width=\linewidth]{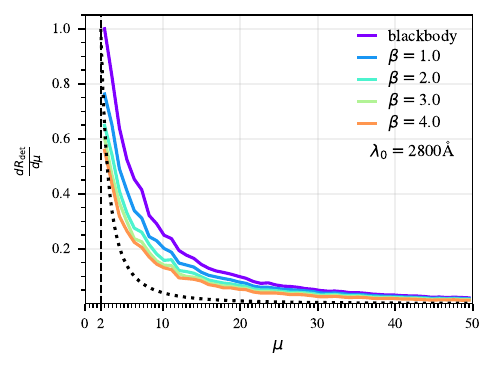} 
    \end{subfigure}
    \caption{The distribution of detected sources as a function of magnification, $\mu$, for a range of different UV suppression models using a galaxy-fold lens model. The detections are normalised so that a value of 1 corresponds to the peak number density of detections for a pure blackbody model. The $\mu > 2$ condition for strong lensing is represented by the dashed line. The dotted curve represents the $\mu^{-2}$ relationship for the probability of a source being lensed. Left: The effect of changing $\lambda_{0}$ for a fixed $\beta = 1.0$. Right: The effect of changing $\beta$ for a fixed $\lambda_{0} = 2800$\r{A}.}
    \label{fig: galaxy_magnification_histogram}
\end{figure*}


\begin{figure*}
    \centering
    \begin{subfigure}[t]{0.49\textwidth}
        \centering
        \hspace{-0.8cm}
        \includegraphics[width=\linewidth]{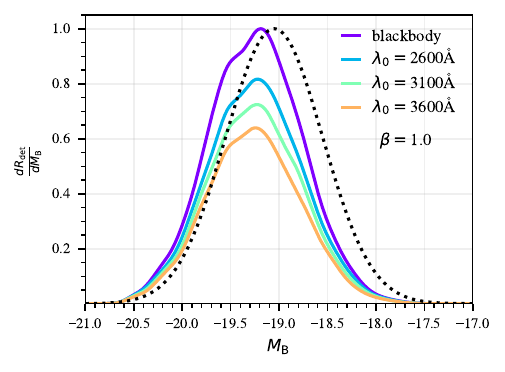} 
    \end{subfigure}
    \begin{subfigure}[t]{0.49\textwidth}
        \centering
        \hspace{-1.2cm}
        \includegraphics[width=\linewidth]{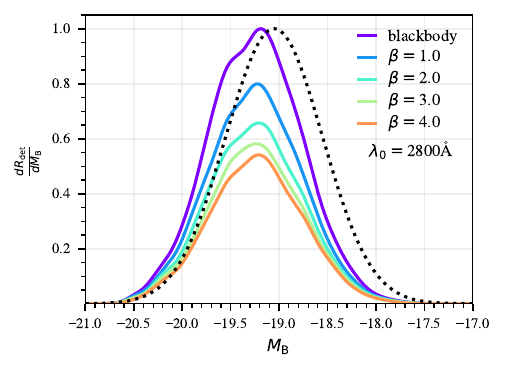} 
    \end{subfigure}
    \caption{The distribution of detected sources as a function ofmthe peak B-band absolute magnitude of the source, $M_{\rm B}$, for a range of different UV suppression models using a galaxy-fold lens model. The detections are normalised so that a value of 1 corresponds to the peak number density of detections for a pure blackbody model. The dotted curve represents the Gaussian distribution for $M_{\rm B}$ used to generate a sample of Type IIn SNe in our simulation, with a mean of -19.05 and a standard deviation of 0.5. Left: The effect of changing $\lambda_{0}$ for a fixed $\beta = 1.0$. Right: The effect of changing $\beta$ for a fixed $\lambda_{0} = 2800$\r{A}.}
    \label{fig: galaxy_magnitude_histogram}
\end{figure*}



\begin{figure*}
    \centering
    \begin{subfigure}[t]{0.49\textwidth}
        \centering
        \hspace{-0.8cm}
        \includegraphics[width=\linewidth]{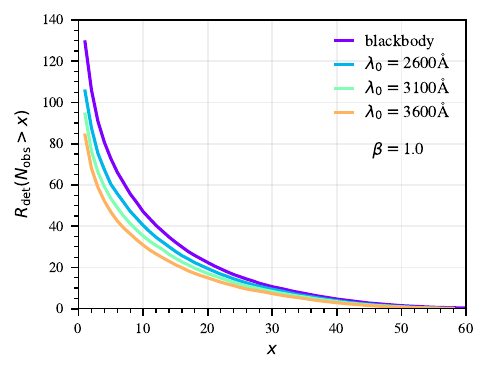} 
    \end{subfigure}
    \begin{subfigure}[t]{0.49\textwidth}
        \centering
        \hspace{-1.2cm}
        \includegraphics[width=\linewidth]{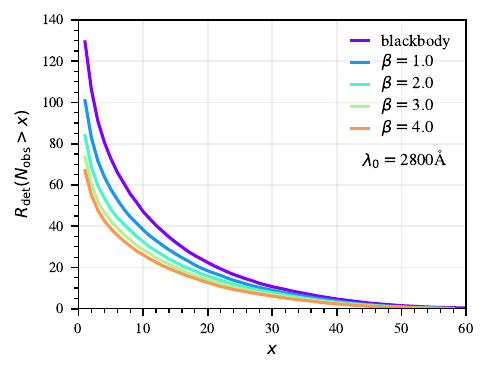} 
    \end{subfigure}
    \caption{Annual expected number of LSST lensed SNIIn detections with $N_{\textrm{obs}}$ total observations  above a threshold $x$. Detections are shown for a range of different UV suppression models, using the galaxy-fold lens model. Left: The effect of changing $\lambda_{0}$ for a fixed $\beta = 1.0$. Right: The effect of changing $\beta$ for a fixed $\lambda_{0} = 2800$\r{A}.}
    \label{fig: galaxy_cumulative_histogram}
\end{figure*}


\subsection{Detection Distributions}\label{sec:detection_distributions}

For each lens model, we determine how the number of detections varies across different values of redshift, magnification and magnitude, across our range of UV suppression models. For images produced by fold caustics in galaxy-scale lenses, these distributions are shown in Figures \ref{fig: galaxy_redshift_histogram}, \ref{fig: galaxy_magnification_histogram} and \ref{fig: galaxy_magnitude_histogram} respectively. To compare between models, the distributions are normalised so that the peak density of the blackbody distribution is equal to 1. In all cases, the number of detections is more sensitive to changing $\beta$ for a fixed $\lambda_0$ than for changing $\lambda_0$ for a fixed $\beta$, across the scales over which we sample. However, it can be seen that the overall effect of increasing each parameter on the amplitude and shape of the distributions is similar. For simplicity, we only show the distributions for the galaxy-scale fold caustic lens model, however the trends described hold for every lens model, with no noticeable differences.

For the distribution with respect to redshift, seen in Figure \ref{fig: galaxy_redshift_histogram}, the blackbody model peaks close to the peak of the star-formation rate density (SFRD) evolution predicted by \cite{madau_2014}, which is used to inform the cosmic rate evolution of Type IIn supernovae (see Section \ref{sec:redshift}). This can be attributed to the optimal scenario in which this is the redshift: a) where the number density of Type IIn SNe peaks, b) where the spectral flux peak is redshifted into the most sensitive LSST bands, and c) where a low magnification ($\mu \sim 2$) is still sufficient to render a source observable. Figure \ref{fig: redshift_magnification}, a plot showing the 90\% and 99\% completeness contours for detections as a joint distribution of redshift and magnification, illustrates these criteria. It shows that, for both an evolving and non evolving source population, a magnification of $\mu \sim 2$ is sufficient to detect lensed Type IIn SNe up to $z > 2.5$, depending on the strength of the UV suppression. In terms of spectra, the non-evolving source contour illustrates that detections up to $z=4$ are plausible, however in the evolving case, the bulk of the detections are at a lower redshift corresponding to the $z \sim 2$ peak of the SFRD. The detection peak at $z \sim 2$ is much steeper than the peak of the SFRD, however, as seen in Figure \ref{fig: galaxy_redshift_histogram}. At the low redshift tail, the detection density drops rapidly relative to the SFRD as the lensing optical depth decreases sharply. Here, the suppression has no effect on the detection rates as the wavelengths in which the detections are made are unaffected. The effect of the suppression starts to take hold strongly above $z = 1.2$, where the suppressed spectrum shifts to the wavelength range of the sensitive LSST bandpasses. The redshift of the detection peak also decreases for increasing suppression, as the ``effective peak'' of the spectral flux density increases in wavelength (see Figure \ref{fig:spectra}). Beyond the $1.5 < z < 2$ peak of the different UV models, the number density again drops sharply relative to the SFRD as the sources are too faint to detect without considerable magnification. Given this effect, UV suppression disproportionately affects the capability of LSST to detect high redshift sources, and decreases the maximum redshift to which we are likely to observe a lensed Type IIn SN during the 10 year survey lifetime.

Regarding magnification, the distribution peaks for a magnification of $\mu = 2$, for all UV suppression models, and has the same shape, with only the amplitude changing. This is due to the fact that low magnification lensing events are sufficient to detect sources that make up the majority of the redshift and absolute magnitude distributions. In every instance, the distribution falls less steeply than the $\mu^{-2}$ relationship obtained from the optical depth (Equation \ref{eqn:tau}). This reflects that a source has a greater chance of being bright enough to be detected as magnification increases.

As expected for the magnitude distribution, we obtain a skewed Gaussian distribution. Compared to the Gaussian distribution for the peak rest-frame B-band absolute magnitudes of the SNe in our sample (see Table \ref{tab:parameters}), the detection distribution has a slightly brighter peak at -19.20, and is skewed so that there is a longer bright tail and a shorter faint tail. This shape is due to the simple fact that brighter events are more likely to be detected. The small size of the shift is indicative that a Type SNIIn with a brightness close to the means is relatively likely to be detected with low magnification at $z=2$, where the cosmic rate of Type IIn SNe peaks. There is no major difference in the shapes for each of the UV suppression models since the magnitude acts as a scale factor in the amplitude of the spectral flux, and so has no correlation with the suppression parameters.

\subsection{Sensitivity to Number of Observations}\label{sec:observation_quality}

One way of quantifying the quality of data obtained for a detected SN is by the number of times it is observed, $N_{\textrm{obs}}$. While this simple metric does not account for whether the observations are taken at key moments in the evolution of the SN, such as observations around the lightcurve peak, it gives a sense of how well sampled the lightcurve is. Figure \ref{fig: galaxy_cumulative_histogram} shows the number of lensed IIn detections per year with $N_{\textrm{obs}} > x$. Promisingly, the yearly detection rate of well sampled lensed Type IIn SNe is still high for moderate suppression models. When considering a SLSNe-like suppression strength of $\lambda_0 \sim 2800$ and $0.8 < \beta < 3.0$ we obtain $\sim$ 30 detections per year with more than 10 data points, $\sim$ 20 with more than 20 data points, and $\sim$ 10 with more than 30 data points.

\begin{figure*}
    \centering
    \begin{subfigure}[t]{0.49\textwidth}
        \centering
        \hspace{-0.8cm}
        \includegraphics[width=\linewidth]{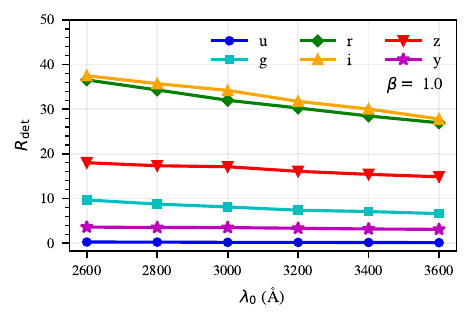} 
    \end{subfigure}
    \hfill
    \begin{subfigure}[t]{0.49\textwidth}
        \centering
        \hspace{-1cm}
        \includegraphics[width=\linewidth]{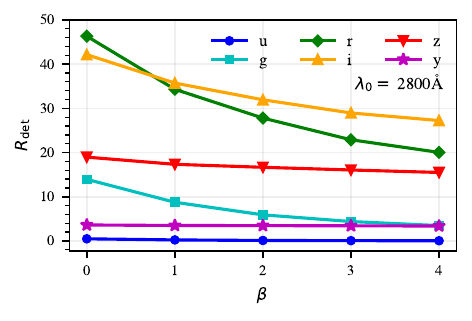} 
    \end{subfigure}
    \caption{Annual expected number of detected lensed Type IIn SNe with at least 5 observations in a given filter. Detections are shown for each of the LSST \textit{ugrizy} filters, for a range of different UV suppression models, using the galaxy-fold lens model. Left: The effect of changing $\lambda_{0}$ for a fixed $\beta = 1.0$. Right: The effect of changing $\beta$ for a fixed $\lambda_{0} = 2800$\r{A}.}
    \label{fig: filter_sensitivity}
\end{figure*}

\subsection{Sensitivity to Filter choice}\label{sec:filter_choice}

To explore how increasing UV suppression has an effect on the LSST filters in which the lensed Type IIn SN is observed, in Figure \ref{fig: filter_sensitivity} we plot the number of lensed SNe detected with at least 5 observations in a given filter band for a range of UV suppression models. As expected, the \textit{g r} and \textit{i} bands are those most affected as they correspond to rest frame UV wavelengths for $z \sim 2$. Notably, the most effective filter transitions from the \textit{r} to the \textit{i} band when going from a blackbody ($\beta = 0$) model to a low UV suppression ($\beta = 1$) model. However, the \textit{r}, \textit{i} and \textit{z} bands remain the principle bandpasses for detections of lensed Type IIn SNe, with only the \textit{g} band, previously comparable with the \textit{z} band using a blackbody model, dropping out when UV suppression is incorporated.

\subsection{Spectroscopic and Time-Delay Gold Sample}\label{sec:spec_detections}

\begin{figure*}
    \textbf{\large Spectroscopic Gold Sample Yearly Detection Rates}\par\medskip
    \centering
        \begin{subfigure}[t]{0.49\textwidth}
        \centering
        \includegraphics[width=\linewidth]{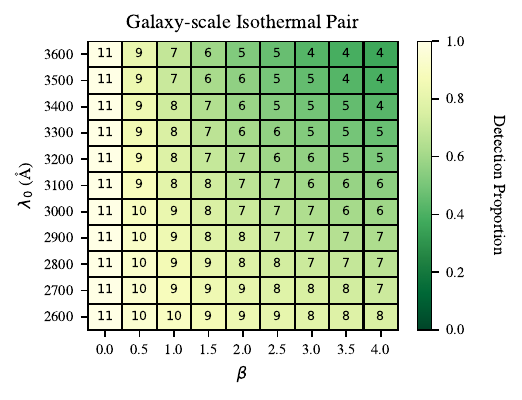} 
    \end{subfigure}
    \hfill
    \begin{subfigure}[t]{0.49\textwidth}
        \centering
        \includegraphics[width=\linewidth]{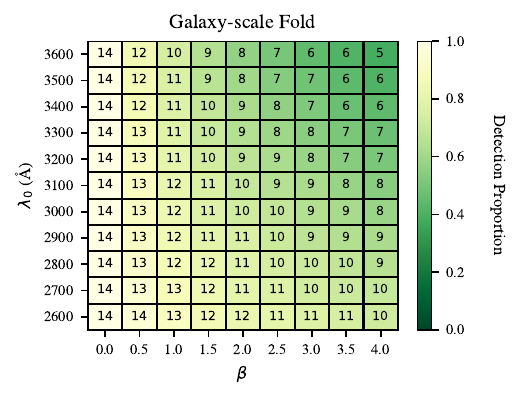} 
    \end{subfigure} \\
    \begin{subfigure}[t]{0.49\textwidth}
        \centering
        \includegraphics[width=\linewidth]{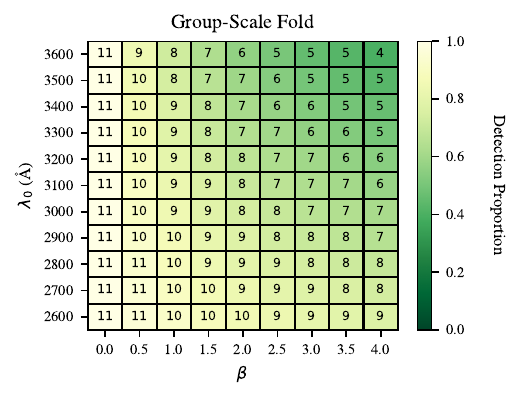} 
    \end{subfigure}
        \hfill
    \begin{subfigure}[t]{0.49\textwidth}
        \centering
        \includegraphics[width=\linewidth]{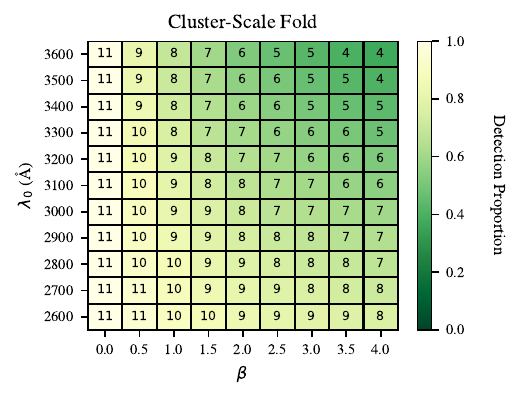} 
    \end{subfigure}
    \caption{The total yearly LSST detection rate of lensed Type IIn SNe that meet the spectroscopic gold sample criteria for a range of UV suppression models. Each square in the grid corresponds to a different combination of power law index, $\beta$, and cutoff wavelength, $\lambda_{0}$. The left hand column corresponds to $\beta = 0$, i.e. a pure blackbody model. The colour bar represents the fraction of detections for each model relative to the spectroscopic gold-sample detections obtained using a pure blackbody model for the given lens model. The four panels each show results for a different lens archetype. From the top-left going clockwise, the lens models represent images formed by: galaxy-scale isothermal lenses, galaxy-scale fold caustics, cluster-scale fold caustics and group-scale fold caustics. These models are described in Section \ref{sec:optical_depth}.}
    \label{fig:detection_heatmaps_green}
\end{figure*}

The total yearly detections of strongly lensed Type IIn SNe that are part of our spectroscopic gold sample, i.e. that meet criteria (i) and (ii) in Section \ref{sec:gold sample}, for a range of different UV suppression models are summarised in Figure \ref{fig:detection_heatmaps_green}. Approximately 10-20\% of the total detections for each UV suppression model meet the criteria for this sample, consisting of $\sim$10 spectroscopically useful lensed Type IIn SN detections per year with LSST. Slightly more detections are obtained for the galaxy-scale fold model as these produce more events with unresolved lightcurves and therefore the overall observed peak brightness occurs later than the peak brightness of the first image. Therefore these surplus detections are less indicative of the desired properties of a spectroscopic gold sample.

Regarding the time-delay gold sample, i.e. lensed Type IIn SNe that meet criteria (i), (ii) and (iii) in Section \ref{sec:gold sample}, the sample size depends strongly on the lens catastrophe responsible for the production of multiple images. For images produced by galaxy-scale isothermal lenses and cluster-scale fold caustics, due to the long time delays and sufficiently large image separations expected for their typical Einstein radii ($\theta_{\rm E} = 1''$ and $\theta_{\rm E} = 10''$ respectively), almost all of SNe that meet the criteria for the spectroscopic gold-sample have sufficient time delays for the time-delay gold sample also. However, for smaller Einstein radii, the time delays and image separations become too small. These assertions are primarily influenced by our choice of a single representative value for the lens parameters for each of our lens archetypes. This is discussed further in Section \ref{sec:cosmo_prospects}.


\section{Discussion}\label{sec:discussion}

The predicted detection rates of strongly lensed Type IIn SNe in this study depend on a range of assumptions. In this section we discuss the effects of these assumptions on our results and how they may be better constrained in the future. In addition we discuss the astrophysical and cosmological prospects for our predicted population of lensed  Type IIn SNe detected by LSST.

\subsection{Detection vs Identification}

In this study, we use a simple detection criterion that classes a SN as detected if it is observed by LSST in 5 or more separate epochs. Most studies that predict the number of lensed SN discoveries generally take into account two detection methods, which intrinsically incorporate identification of the SN as being strongly lensed. The first is the \textit{image multiplicity method}, where a lensed SN is confirmed by having multiple resolved images of the SN \citep{Oguri2010}. The second is the \textit{magnification method}, where a SN is deemed lensed if it is significantly brighter than a typical SN at the redshift of the lens \citep{Goldstein_2017}. These detection methods are good indications that an object may be a lensed SN. However, unlike Type Ia SNe, Type IIn SNe have a very wide distribution of peak magnitudes. This distribution is also poorly constrained and varies from sample to sample. For instance, \citet{goldstein_2019} use a distribution for the peak B-band absolute magnitude with a mean of $-19.05$ and a scatter of $0.5$, whereas \citet{wojtak_2019} use a distribution with a mean of $-18.5$ and a scatter of $1.0$ \citep[adapted from][]{rson_2014}. There is also a suggestion that the distribution of IIn magnitudes extends into the region usually classified as SLSNe. Therefore, strongly lensed SNe from the low-luminosity tail of this SNe will have a luminosity degenerate with that of a bright non-lensed SNIIn, making it difficult to apply the magnification method to identify lensed Type IIn SNe. While we could have incorporated the image multiplicity method, we do not vary our parameters for each of our lens-archetypes, and so the distribution of Einstein radii and time delays would not reflect a true lens population. However, we assert that UV suppression is likely to skew the time delay distribution to shorter time delays as the suppression leads to fewer higher redshift detections, potentially making identification more difficult. We argue that, other than this effect, the proportional effect of UV suppression should be agnostic to the detection and identification criteria used, provided the magnification method uses absolute magnitude measurements in an optical band. Therefore the trends in the relative decreases in detections between the blackbody and UV suppression models should hold for different methods of determining detection and identification.

Regarding the identification of a SN as a Type IIn SN, care must be taken. From photometry alone, the lightcurve of some Type IIn SNe could lead them to be classified as SNIIL \citep[e.g.][]{Fassia_2000, Kangas_2015, Benetti_2016}. Even with spectroscopy, a Type IIn SN may be misidentified if it only observed  during an epoch where the CSM-interaction is not present. Therefore, long-term spectroscopic monitoring of sources is the only way to accurately classify them as a Type IIn. Promisingly, we have determined from our spectroscopic gold sample (see Section \ref{sec:gold sample}) that at least $\sim$ 10 lensed Type IIn SNe per year can confidently be classified in this manner, and we expect more to be classified with less frequent spectroscopy.

\subsection{Cosmic Rate of Type IIn SNe}

 We model the cosmic rate of Type IIn SNe to mirror the star formation rate density evolution with redshift. In doing so, we assume that there is zero time delay between star formation and SN, motivated by the relatively short lifespan of giant stars comprising expected SNIIn progenitors. However, the lifespan of these progenitors can be extended due to effects associated with binary formation channels, such as mass transfer and mergers \citep{zapartas_2017}. Furthermore, the prevalence of these effects at different cosmic epochs may vary, leading to a time delay distribution that changes with redshift. \citeauthor{zapartas_2017} find time delays that range from  $5$Myr $<$ t $< 250$Myr, which would have an effect when mapping from the star formation rate density to the SNIIn rate density, especially around the $z = 2$ peak. This would shift the overall redshift distribution of detected sources to lower redshift. The size of the shift depends on the true time delay distribution, including the potential for the distribution to change with redshift as certain formation channels become more or less favoured. 
 
 Similarly, we assume that the fraction of Type IIn SNe relative to other CCSNe is constant across time, fixed at the detection fraction obtained for local SNIIn observations in \cite{cold_2023}. In reality, as discussed by \citeauthor{cold_2023} it is likely that the relative rates of Type IIn SNe to CCSNe will evolve with redshift, due to different stellar environments at high redshift, and this evolution will be reflected in the overall detection rates. With our expected sample LSST sample of lensed SNIIn detections spanning a large redshift range, we can begin to determine how the relative SN rates evolve with redshift, and thereby indirectly probe the stellar environment, SN progenitor systems and their formation channels at high redshift.

\subsection{Type IIn Supernova Heterogeneity}
\label{subsec:heterogeneity}

Our choice to use a single template SNIIn lightcurve is motivated by the decision to make comparisons to previous studies that use the same template, and to focus our analysis solely on the effect of UV suppression on the detection rates. In reality, Type IIn SNe exhibit an enormous amount of heterogeneity. They span about two orders of magnitude in peak luminosity, an order of magnitude in timescale, and three orders of magnitude in total radiated energy, with strong correlations in luminosity and timescale \citep{Hiramatsu_2024}. While we vary the peak luminosity, this distribution is poorly constrained, with values obtained from different samples varying widely. For instance, the distribution used in this work and in \cite{goldstein_2019} has a mean peak B-band absolute magnitude of -19.05 and a scatter of 0.5, whereas \cite{wojtak_2019} use a distribution with a mean of -18.5 and a scatter of 1.0, obtained from \cite{rson_2014}. \citet{Hiramatsu_2024} obtain a brighter and broader range of magnitudes, with a g-band peak at -19.21 mag and a scatter of $\sim$ 1.3 mag. This discrepancy exists due to a variety of reasons. 

Firstly, there are a wide range of progenitor channels for Type IIn SNe \citep[see][and references therein]{Fraser_2020}. \citet{Hiramatsu_2024} argue that due to a clear bimodality in the luminosity-timescale correlations of Type IIn SNe, with luminous-slow and faint-fast groups, there are likely two distinct dominant progenitor channels each responsible for one of these groups. 

There is also debate as to where the brightness distributions of Type IIn SNe end. The lower luminosity tail likely falls off too rapidly due to Malmquist bias and the higher luminosity tail is often truncated at $M_{\rm R} = -21$, beyond which the SNe are classed as a separate class of superluminous Type II SNe. Some advocate that this classification should be abandoned; that these SNe are just the high luminosity tail of the SNIIn population, and have the same properties \citep[e.g.][]{Hiramatsu_2024}. 

Furthermore, this magnitude distribution may also vary with redshift, since the stellar properties at high-redshift, in particular the density and composition of the CSM, are potentially different to those in the local universe. 

In order to perform a comprehensive analysis of lensed SNIIn discovery rates by LSST, a simulated sample of SNe must encapsulate this diversity. On the other hand, the sample of lensed SNIIn detections expected from LSST can help to further understand and potentially model this heterogeneity, and how the various properties change with redshift. Care must be taken to determine which aspects of the observed population distributions are reflective of the true source population and which are due to observational biases.

\subsection{UV Suppression Assumptions}

For each suppression model, we have used a single value of $\beta$ and $\lambda_0$. In reality, there is likely to be some variance in the distributions of these values, as is seen in the small sample of SLSNe in \cite{yan_2018}. However, due to the smooth transitions in the distributions of all the observables as $\beta$ and $\lambda_0$ each increase, these models can be interpreted as Gaussian distributions centred at the single values of $\beta$ and $\lambda_0$. Unfortunately, comparing the distribution of detections with real LSST lensed IIn SN detections can similarly only yield an inference for the mean for the distributions of $\beta$ and $\lambda_0$. To investigate the true spread of the UV suppression, we must fit UV models to spectra of observed lensed Type IIn SNe. This requires a coordinated effort to spectroscopically follow up as many candidate lensed Type IIn SNe as possible. The prevalence of bright detections around $z = 2$ offers distinct advantages since the source frame UV spectra is redshifted into optical bands, and many of the SNe are optically bright for months. Therefore, LSST will provide a large sample of lensed Type IIn SNe that can feasibly be followed up with optical spectroscopy.

Additionally, the distribution of $\beta$ and $\lambda_0$ is expected to evolve with redshift, as the density and properties of the CSM, and properties of the progenitor systems, both evolve. Once more, spectroscopic follow up is required to determine how the distribution of the suppression evolves with redshift. However, without spectroscopy, deviations of the true redshift distribution of detections with our predicted distribution may hint at how the average UV suppression model changes with redshift.

Thirdly, the processes that cause the UV suppression are expected to vary in time. Many CSM-interacting SNe only exhibit SNIIn-like features early in their evolution \citep[e.g.][]{Fassia_2001, Yaron_2017} or late in their evolution \citep[e.g.][]{Chugai_2006, Milisavljevic_2015,Benetti_2018}. If there is a relationship between the strength of the UV suppression and the properties of the CSM, then we could feasibly expect the strength of the UV suppression to vary as the SN goes through phases of more or less CSM interaction. Therefore, in an ideal population study, a UV suppression model should encapsulate variation over time.

\subsection{High cadence spectroscopy of Type IIn SNe}

The key to learning more about the interaction between SN ejecta and CSM, the strength of the UV suppression, and how it varies over time, is to obtain frequent spectroscopic observations of the SN from pre-peak and throughout its evolution. SNe in our sample that exhibit the properties that render them capable of such observations form our spectroscopic gold sample (see Section \ref{sec:gold sample}). Using these criteria, we obtain between 8-10 lensed Type IIn SNe detected per year by LSST for which follow-up high cadence spectroscopic observations are feasible, given a similar UV deficit to that seen in SLSNe ($1.0 < \beta < 3.0$, $\lambda_0 = 2800$\r{A}). This is a promising sample size to begin to understand the nature of CSM interactions in Type IIn SNe at a range of redshifts. As mentioned before, this will also allow us to investigate the cause and effects of aspherical and time-varying CSM, including the processes that drive the winds that sweep up the CSM in later epochs. Another aspect of our spectroscopic gold sample is the requirement for at least 5 epochs of photometry pre-peak. As shown recently in \cite{Hinds_2025}, the existence and the extent of a dense CSM, as well as other pre-explosion progenitor properties, can be inferred from SN rise times, which are obtained from pre-peak photometry. 

\subsection{Cosmological prospects of Type IIn SNe}\label{sec:cosmo_prospects}

In order to obtain a useful time-delay measurement with lensed SNe, the lightcurve must be sufficiently well-sampled in order to accurately fit a model to it. \cite{arendse_2024} find that, for lensed Type Ia SNe, the LSST cadence leads to sparse sampling and unsuccessful light curve fits in almost all cases, with less than 2$\%$ of observed SNe allowing for a time-delay measurement with better than 5$\%$ accuracy. Due to the heterogeneity exhibited in Type IIn SN lightcurves and a lack of robust lightcurve models, we expect the proportion of lensed Type IIn for which we can calculate accurate time-delays with LSST observations alone to be even lower. Therefore, a more sensible strategy is to try to identify lensed SNe early in their evolution so that we can conduct timely follow-up observations. 

Using the criteria outlined in Section \ref{sec:gold sample} for the time-delay gold sample, \citet{arendse_2024} find that $\sim 25\%$ of lensed Type Ia SNe detected by LSST are part of this gold sample. This equates to $\sim 10$ gold sample lensed Type IIn SNe if a similar fraction of the population meets these criteria. By computing this time-delay gold sample in the same manner, we obtain values that are consistent with this estimate. Qualitatively speaking, however, it is likely that there will be more time-delay gold sample lensed Type IIn SNe than lensed Type Ia SNe since lensed Type IIn SNe are more likely to meet the three criteria required. For criterion (i): the typical rise time to peak for Type IIn SNe is $\sim 39$ days \citep{Ransome_2024} which is double that of $\sim 19$ days for Type Ia SNe \citep{Miller_2020}. For criterion (ii): lensed Type IIn SNe are bright in the i-band at the redshift they are most commonly detected. For criterion (iii): the expected time delays will be longer for Type IIn SNe as they are more commonly detected at a higher redshift than Type Ia SNe. Our result is affected by choosing a fixed lightcurve model with a shorter rise time to peak than the expected mean rise time of a population of Type IIn SNe, which is primarily dictated by the amount of CSM and can vary significantly. Therefore, we cut more samples for criterion (i) than the true population would dictate. Our result is also affected by choosing a fixed value of Einstein radius for each lens archetype, leading to distributions of time-delay and image separation that do not fully encapsulate the diversity of lenses.

A key hurdle to obtaining accurate values for cosmological parameters from time-delays is the so-called "mass-sheet degeneracy" (MSD), where observed image positions and time delays can be recreated by many different mass models \citep{Falco_1985, Schneider_2013}. Type Ia SNe overcome this since they are "standardisable candles" and so provide a direct measurement of magnification, which can be used to help break the MSD, as well as a time delay. For lensed Type IIn SNe (or lensed CCSNe in general), alternative methods of breaking the MSD are required. For instance, obtaining spatially resolved stellar velocity dispersion data of the lensing galaxy (or galaxies for group and cluster-scale lenses) allows us to break the MSD by constraining its mass profile independently \citep{birrer_2021, shajib_2023}. Alternatively, the mass distribution can be obtained using high-resolution imaging of the lens galaxy (or galaxies) with an assumed mass-to-light ratio alongside a model for dark matter. Both these methods benefit from the transient nature of SNe since follow-up observations of the lens can be conducted once the SN has faded. Another method of modelling the lens involves multi-plane lensing, where multiple images of two or more sources, located at different redshifts, can provide information about the total mass profile of the lens \citep{Soucail_2004, Jullo_2010}. This method is particularly efficient in lens galaxy cluster with a large number of spectroscopically confirmed multiple images \citep{Caminha_2016, Johnson_2016, Acebron_2017, Grillo_2018}, but also galaxy-galaxy lenses with multiple source planes, where modelling the lenses is more straightforward \citep[see][and references therein]{Collett_2020}.

Putting this all together, over the lifespan of LSST, we expect to obtain an extensive sample of lensed Type IIn SNe for which we can break the MSD and obtain accurate time-delay measurements through follow-up observations. While Type Ia SNe can help break the MSD directly, for Type IIn SNe, the longer period over which they are bright, and the on average longer arrival time differences of their images, provide a strong advantage when it comes to getting follow up data and measuring these arrival time differences. Combining this advantage with their abundance, it is likely that LSST will provide a sample of cosmologically useful lensed Type IIn SNe that could provide a complimentary statistical estimate of $H_0$ and other cosmological parameters.

\section{Conclusion}\label{sec:conclusion}

In this work we have quantified the effect of UV suppression on the detection rates of a simulated population of strongly lensed Type IIn SNe by LSST. Using a blackbody model, we expect to detect $\sim$70 lensed Type IIn SNe per year with LSST. By modelling a similar UV deficit to that seen in SLSNe ($1.0 < \beta < 3.0$, $\lambda_0 = 2800$\r{A}), we recover 60-80\% of the detections obtained using a pure blackbody model. Even with an extreme, and likely unphysical, suppression model ($\beta = 4.0$, $\lambda_0 = 3600$\r{A}), 30\% of the blackbody detections are recovered, which constitutes $\sim$20 detections per year. We therefore conclude that the effect of UV suppression is not enough to extinguish the expected strongly lensed Type SNIIn detections by LSST. Of these detections, many are well sampled, with $\sim$20 detections per year observed for at least 20 epochs, and are optically bright, with $\sim$30 detections per year observed for at least 5 epochs in the i-band. Imposing the requirements of 5 epochs of pre-peak detections across two filters and a SN which is sufficiently bright for spectroscopic follow-up, we obtain $\sim$ 10 detections per year that make up our spectroscopic gold sample, for a SLSN-like suppression model. Therefore the prospects of using lensed Type IIn SNe for the study of CSM-interaction, the terminal stages of stars, progenitor diversity, and as cosmological probes, are bright. With optical spectroscopic follow up possible for many of the detections, we expect to learn more about their UV characteristics and use them to inform future population studies. Given their brightness, expected longer arrival time difference of their images, and abundance relative to Type Ia SNe, it is also likely the LSST will yield a competitive sample of lensed Type IIn SNe useful for cosmological inferences and measurements of $H_0$, provided that the MSD can be addressed via alternative means.

\section*{Acknowledgements}

This paper has undergone internal review in the Rubin LSST Strong Lensing Science Collaboration (SLSC). The authors would like to thank Satadru Bag for their insightful comments and review.
AIPP thanks Ariel Goobar, Bill Chaplain, Maurice Meus, Benjamin Gompertz, and colleagues at the University of Birmingham for the stimulating discussions that helped to shape this paper.

Author contributions are listed below.

AIPP: conceptualization, methodology, software (lensed SN simulations), formal analysis, writing (original draft; review and editing),
visualization;

GPS: conceptualization, methodology, validation, writing (review and editing), supervision;

MN: conceptualization, methodology, writing (review);

NA: methodology, software (lensed SN simulations), validation, writing (review and editing);

DR: conceptualization, writing (review);

SD: writing (review).

AIPP acknowledges a PhD studentship from the Science and Technology Facilities Council and the University of Birmingham. GPS acknowledges support from The Royal Society, the Leverhulme Trust, and the Science and Technology Facilities Council (grant number ST/X001296/1).
MN is supported by the European Research Council (ERC) under the European Union’s Horizon 2020 research and innovation programme (grant agreement No.~948381) and by UK Space Agency Grant No.~ST/Y000692/1. NA is supported by the research project grant ‘Understanding the Dynamic Universe’ funded by the Knut and Alice Wallenberg Foundation under Dnr KAW 2018.0067. SD is supported by UK Research and Innovation (UKRI) under the UK government’s Horizon Europe funding Guarantee EP/Z000475/1 and acknowledges a bye-fellowship at Lucy Cavendish College.

\medskip
\textit{Software:} Astropy \citep{Astropy2013, Astropy2018, Astropy2022},
Jupyter \citep{JupyterNotebook},
Matplotlib \citep{Matplotlib2007, Matplotlib2020},
NumPy \citep{Numpy202},
Pandas \citep{Pandas2010, Pandas2023},
Pickle \citep{Pickle2020},
SciPy \citep{Scipy2020},
SQLite \citep{sqlite2020hipp},
SNcosmo \citep{barbary_2024},
OpSimSummary \citep{Biswas_2020},
tqdm \citep{tqdm_2024}.

\section*{Data Availability}

The data presented in this article are available upon reasonable requests to the lead author.



\bibliographystyle{mnras}
\bibliography{main} 
\bsp	
\label{lastpage}
\end{document}